\documentclass[structabstract]{aa}  
\usepackage{graphicx}
\usepackage{txfonts}
\usepackage{natbib}
\begin{document}

\title{Multiplicity and disks within the high-mass core NGC7538IRS1}

\subtitle{Resolving cm line and continuum emission at $\sim
  0.06''\times 0.05''$ resolution}

   \author{H.~Beuther
          \inst{1}
          \and
          H.~Linz
          \inst{1}
          \and
          Th.~Henning
          \inst{1}
          \and
          S.~Feng
          \inst{2}
          \and
          R.~Teague
          \inst{1}
}
   \institute{$^1$ Max Planck Institute for Astronomy, K\"onigstuhl 17,
              69117 Heidelberg, Germany, \email{beuther@mpia.de}\\
              $^2$  Max Planck Institute for Extraterrestrial Physics, Giessenbachstrasse 1, 85748 Garching
}

   \date{Version of \today}

\abstract
{High-mass stars have a high degree of multiplicity and most likely
  form via disk accretion processes. The detailed physics of the
  binary and disk formation are still poorly constrained.}
{We seek to resolve the central substructures of the prototypical high-mass
  star-forming region NGC7538IRS1 at the highest possible spatial
  resolution line and continuum emission to investigate the
  protostellar environment and kinematics.}
{Using the Karl G.~Jansky Very Large Array (VLA) in its most extended
  configuration at $\sim$24\,GHz has allowed us to study the NH$_3$ and
  thermal CH$_3$OH emission and absorption as well as the cm continuum
  emission at an unprecedented spatial resolution of $0.06''\times
  0.05''$, corresponding to a linear resolution of $\sim$150\,AU at a
  distance of 2.7\,kpc.}
{A comparison of these new cm continuum data with previous VLA
  observations from 23\,yrs ago reveals no recognizable proper
  motions. If the emission were caused by a protostellar jet, proper
  motion signatures should have been easily identified. In combination
  with the high spectral indices $S\propto \nu^{\alpha}$ ($\alpha$
  between 1 and 2), this allows us to conclude that the continuum emission
  is from two hypercompact H{\sc ii} regions separated in projection
  by about 430\,AU. The NH$_3$ spectral line data reveal a common
  rotating envelope indicating a bound high-mass binary system. In
  addition to this, the thermal CH$_3$OH data show two separate
  velocity gradients across the two hypercompact H{\sc ii}
  regions. This indicates two disk-like structures within the same
  rotating circumbinary envelope. Disk and envelope structures are
  inclined by $\sim$33\,$^{\circ}$, which can be explained by initially
  varying angular momentum distributions within the natal, turbulent
  cloud.}
{Studying high-mass star formation at sub-0.1$''$ resolution allows us
  to isolate multiple sources as well as to separate circumbinary from
  disk-like rotating structures. These data show also the
  limitations in molecular line studies in investigating the disk
  kinematics when the central source is already ionizing a
  hypercompact H{\sc ii} region. Recombination line studies will be
  required for sources such as NGC7538IRS1 to investigate the gas
  kinematics even closer to the protostars.}  

\titlerunning{Multiplicity and disks within the high-mass core NGC7538IRS1}

\maketitle

\section{Introduction}
\label{intro}

The formation processes leading to the most massive stars are still
puzzling in many ways. While there is a clear consensus that high-mass
stars shape the interstellar medium, whole clusters, and even entire
galaxies, questions related to the physical processes at early
evolutionary stages, in particular associated with the fragmentation
processes of young, dense, and rotating cores, formation of
multiple entities, and embedded accretion disks, are still
poorly explored (e.g.,
\citealt{beuther2006b,zinnecker2007,tan2014,beltran2016}).

How do massive accretion disks form and what are their properties?
These are central questions for high-mass star formation research
(e.g.,
\citealt{henning2000,kratter2006,beuther2009c,cesaroni2007,vaidya2009,kraus2010,kraus2017,beltran2011,ilee2013,boley2013,boley2016,sanchez2014,johnston2015}).
The main indirect evidence stems from observations of massive and
collimated outflows that are qualitatively similar to low-mass jets (e.g.,
\citealt{beuther2002d,zhang2005,arce2006,lopez2009,duarte2013}).  Such
jet-like outflows are best explained via magneto-centrifugal
acceleration from an accretion disk and subsequent
Lorentz collimation. Radiation (M)HD simulations of massive collapsing
cores produce accretion disks as well (e.g.,
\citealt{yorke2002,krumholz2009,kuiper2010,kuiper2011,kuiper2013,peters2010b,commercon2011}).
Are massive disks similar to their low-mass counterparts, hence
dominated by the protostar and Keplerian rotation or are they
self-gravitating non-Keplerian entities (e.g.,
\citealt{sanchez-monge2013,cesaroni2007})? The answer may be a small
inner Keplerian accretion disk that is fed from a larger-scale
non-Keplerian structure (toroid or pseudo-disk).  This picture is
supported by analytic and numeric models with Keplerian disks growing
with time from the infalling rotating structure (e.g.,
\citealt{stahler2005,kuiper2011}).  The transition from molecular to
ionized infall is an additional important characteristic
(e.g., \citealt{keto2002a,keto2003,klaassen2009,klaassen2013}).  While
indirect evidence for massive disks is very strong, direct
observations are still sparse (see AFGL4176 as one of the best recent
examples; \citealt{johnston2015}, of G11.92-0.61 by
\citealt{ilee2016}).  This discrepancy is mainly due to the clustered
mode of massive star formation at large distances.  High spatial
resolution is crucial to disentangle these structures.

\begin{figure*}[htb]
  \includegraphics[width=0.99\textwidth]{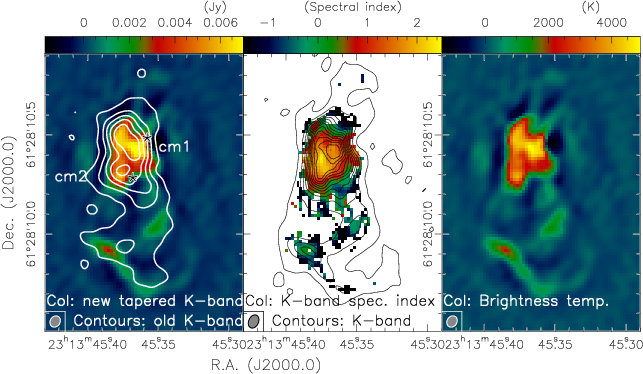}
  \caption{Centimeter continuum emission from NGC7538IRS1. The {\it
      left} panel shows in color scale the new 1.2\,cm continuum data
    imaged using only baselines between 10 and 37\,km, achieving a
    spatial resolution of $0.06''\times 0.05''$. The contours present
    for comparison the old VLA data from 1992 discussed previously in
    \citet{gaume1995}, \citet{sandell2009}, \citet{moscadelli2014},
    and \citet{goddi2015} starting at the $4\sigma$ contour and
    continuing in $8\sigma$ intervals
    ($1\sigma\sim0.05$\,mJy\,beam$^{-1}$). The two stars indicate the
    CH$_3$OH maser positions by \citet{moscadelli2014}; see section
    \ref{cm_cont} for more details. The {\it middle} panel shows in
    color the spectral index map derived from the new full dataset
    with robust weighting $-2$ and the contours show the corresponding
    continuum image using all data at a resolution of $0.07''\times
    0.05''$. The contour levels start at a $4\sigma$ levels of
    0.16\,mJy\,beam$^{-1}$ and continue in coarser 32$\sigma$
    steps. The {\it right} panel again shows the new $0.06''\times
    0.05''$ data, now converted in the Rayleigh-Jeans approximation to
    brightness temperature.}
\label{continuum} 
\end{figure*} 

What are the fragmentation properties of massive gas clumps during the
formation of high-mass stars and their surrounding clusters?
High-mass stars form in clusters with a high degree of multiplicity,
and \citet{chini2012} argue that this multiplicity likely stems from
the formation processes (see also
\citealt{peter2012}). \citet{peter2012} find companion separations
between 400 and several thousand AU, stressing the necessity of high
spatial resolution, Furthermore, \citet{sana2012} infer that multiple
system interactions dominate the evolution of massive stars.
Interferometer studies have revealed that most high-mass star-forming
regions fragment into multiple objects, suggesting that massive
monolithic cores larger than several 1000\,AU are rare, however, the
degree of fragmentation varies (e.g.,
\citealt{cesaroni2005,beuther2007d,beuther2012c,zhang2009,bontemps2010,wang2011,rodon2012,palau2013,sanchez2014,johnston2015}).
Even regions that remain single continuum sources down to arcsec
resolution, mostly fragment on even smaller scales.

\begin{figure}[htb]
  \includegraphics[width=0.49\textwidth]{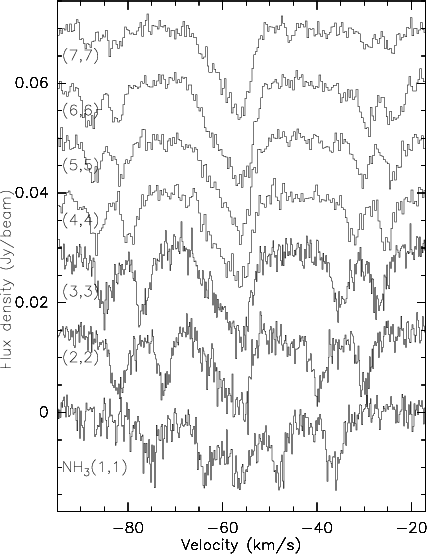}
  \caption{NH$_3$ example spectra extracted toward the northern cm
    continuum peak position. The spectra are shifted of the y-axis for
    presentation purposes. All lines are labeled.}
\label{spectra_nh3} 
\end{figure} 

\begin{figure}[htb]
  \includegraphics[width=0.49\textwidth]{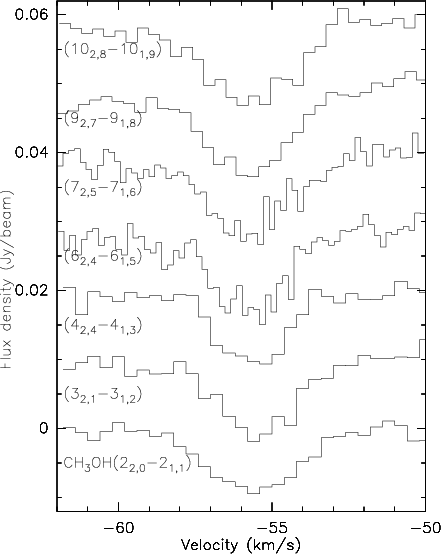}
  \caption{CH$_3$OH example spectra extracted toward the northern cm
    continuum peak position. The spectra are shifted of the y-axis for
    presentation purposes. All lines are labeled.}
\label{spectra_ch3oh} 
\end{figure} 

A particularly revealing example is the famous high-mass star-forming
region NGC7538IRS1. At a distance of $\sim$2.7\,kpc
\citep{moscadelli2009}, the luminosity of the central energy source is
estimated to stem from a 30\,M$_{\odot}$ O6 star (e.g.,
\citealt{willner1976,gaume1995,moscadelli2009}). The strong emission
of this source from near- to mid-infrared wavelengths to the cm- and mm-regime has made it a
famous high-mass protostar for over several decades; a summary of the
literature can be found in \citet{beuther2012c}. The recent $0.2''$
observations at submm wavelengths with the Northern Extended
Millimeter Array (NOEMA) revealed fragmentation of the envelope,
however, these observations did not allow us to identify a Keplerian accretion disk
(\citealt{beuther2013b}, see also high-resolution data by
\citealt{zhu2013}).  Most likely, such a Keplerian structure is
hidden on still smaller scales below 500\,AU (e.g.,
\citealt{krumholz2007a,kuiper2010,kuiper2011}). Furthermore, this
region reveals two cm continuum sources at approximately $0.2''$
separation that may either be two hypercompact H{\sc ii} region or an
ionized jet \citep{gaume1995b,sandell2009,moscadelli2014,goddi2015},
where the association of the potential ionized jet with the molecular
outflow is debated \citep{knez2009,beuther2013b}.

\section{Observations} 
\label{obs}

We observed NGC7538IRS1 on July 21, 2015, during a four hour track
with the Karl G.~Jansky Very Large Array (VLA) in its most extended
A-configuration (baselines extending out to 37\,km). The proposal ID
is 15A-115. With the flexible VLA correlator we covered many spectral
lines and the cm continuum emission in the radio
 K band. Specifically we covered seven NH$_3$ inversion lines, seven
CH$_3$OH lines, and two H$\alpha$ recombination lines. Line
parameters are given in Table \ref{lines}. The following analysis
concentrates on the cm continuum, NH$_3$, and CH$_3$OH
emission. Although the recombination lines are detected, the emission
is comparably weak and is not discussed further here. The
intrinsic spectral resolution for the molecular line data varied
between 15.625 and 31.25\,kHz, corresponding to a velocity resolution
of $\sim$0.19 and $\sim$0.38\,km\,s$^{-1}$ at the given frequencies,
respectively (Table \ref{lines}). Since this region is very strong in
absorption and emission, we reduced almost all lines at the native
correlator resolution. Only the three lowest energy CH$_3$OH lines --
located within a single spectral window -- were reduced separately
with 0.4\,km\,s$^{-1}$ resolution. To create the continuum image, 16
spectral windows with a width of 112\,MHz each between 23.6 and
25.8\,GHz were combined.

The flux and bandpass were calibrated with the two strong quasars 3C48 and
J0319+4130 (also known as 3C84), respectively. The phase and amplitude
gain calibration for these long baselines requires comparatively
  fast switching between the target source and the gain calibrator
  J2339+6010. Our loop typically stayed 1\,min 50\,sec on source and
1\,min 20\,sec on the gain calibrator. We visited the source during
the 4\,h track 58 times assuring an excellent uv coverage. The phase
center of the VLA for our target source NGC7538IRS1 was
R.A.~(J2000.)  23:13:45.36 and Dec.~(J2000.0) +61:28:10.55. The data
calibration was conducted with the VLA pipeline 1.3.1 in CASA
4.2.2. All solutions were carefully checked, and the bandpass for the
NH$_3$(1,1) line was bad during half of the track. Flagging this
second half for this one spectral window and then rerunning the
pipeline gave excellent results.

\begin{figure*}[htb]
  \includegraphics[width=0.99\textwidth]{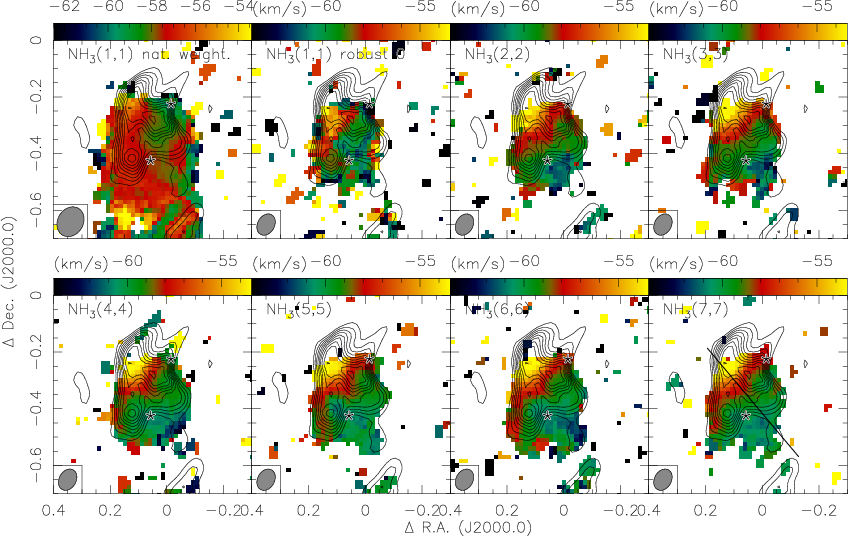}
  \caption{Color-scale presents the 1st moment maps
    (intensity-weighted velocities) of the NH$_3$ inversion
    transitions from (1,1) to (7,7) as indicated in each panel. The first
    two left panels show the data for the NH$_3$(1,1) line with
    different weighting schemes (natural weighting and robust
    weighting, which is a hybrid between natural and uniform
    weighting). The other NH$_3$ lines are always presented with
    robust weighting 0. The 1st moment maps are clipped at a
    $\sim$4$\sigma$ level. The contours show in all panels the 1.2\,cm
    continuum emission in levels of 8\% to 98\% of the peak emission
    (7.4\,mJy\,beam$^{-1}$). The two stars indicate the CH$_3$OH maser
    positions by \citet{moscadelli2014}. The line in the bottom right
    panel indicates the position-velocity cut shown below.}
\label{nh3} 
\end{figure*} 

Further imaging and analysis of the data was also conducted in
CASA. The continuum data were imaged with two different approaches:
once using all data with a robust weighting value of -2, and once
excluding baselines of the inner 10\,km (baselines covered between 10
and 37km) to even improve the spatial resolution. While the normal
robust -2 dataset with all data resulting in a beam of $0.07''\times
0.05''$ (PA $-25^{\circ}$) was better suited for studying the spectral
index, the highest-spatial-resolution image with restricted baseline
range and a spatial resolution of $0.06''\times 0.05''$ (PA
$-32^{\circ}$) was used for morphological comparison. The largest
scales typically recoverable with the VLA at this frequency in the
A-array are $\sim 2.4''$. The 1$\sigma$ rms for both images is
$\sim$0.05\,mJy\,beam$^{-1}$. The molecular line data were all imaged
with a robust weighting scheme and a robust value of 0. Just the
NH$_3$(1,1) data were also imaged in natural weighting (robust value
2) for comparison (see section \ref{molecules}). While the naturally
weighted NH$_3$(1,1) image has a beam of $0.11''\times 0.09''$ (PA
$-31^{\circ}$), the other images with robust weighting 0 have a
synthesized beam of $0.08''\times 0.06''$ (PA varying between
$-28^{\circ}$ and $-30^{\circ}$). The $1\sigma$ rms measured in an
emission- and absorption-free channel varies between 2.1 and
3.9\,mJy\,beam$^{-1}$.

\begin{table}[htb]
\caption{Parameters of main spectral lines}
\begin{tabular}{lrrrr}
\hline
\hline
Line & Freq. & $E_l/k$ & $\Delta \nu$ & $n_{\rm{crit}}^a$ \\
     & (GHz) & (K)     & (kHz)        & ($\frac{10^3}{\rm{cm}^3}$)\\
\hline
NH$_3$(1,1)                 & 23.694506 & 22 & 15.625 & 2 \\
NH$_3$(2,2)                 & 23.722634 & 63 & 15.625 & 2 \\
NH$_3$(3,3)                 & 23.870130 & 122& 15.625 & 2 \\
NH$_3$(4,4)                 & 24.139417 & 199& 31.25  & 2 \\
NH$_3$(5,5)                 & 24.532989 & 294& 31.25  & 2 \\
NH$_3$(6,6)                 & 25.056025 & 407& 31.25  & 2 \\
NH$_3$(7,7)                 & 25.715182 & 537& 31.25  & $^b$ \\
CH$_3$OH($2_{2,0}-2_{1,1}$)   & 24.934382 & 28 & 31.25  & 36 \\
CH$_3$OH($3_{2,1}-3_{1,2}$)   & 24.928715 & 35 & 31.25  & 40 \\
CH$_3$OH($4_{2,4}-4_{1,3}$)   & 24.933468 & 44 & 31.25  & 31 \\
CH$_3$OH($6_{2,4}-6_{1,5}$)   & 25.018123 & 70 & 15.625 & 29 \\
CH$_3$OH($7_{2,5}-7_{1,6}$)   & 25.124872 & 86 & 15.625 & 32 \\
CH$_3$OH($9_{2,7}-9_{1,8}$)   & 25.541398 & 126& 31.25  & 34 \\
CH$_3$OH($10_{2,8}-10_{1,9}$) & 25.878266 & 149& 31.25  & 32 \\
\hline                                          
\hline                                          
\end{tabular}
{~\\
Notes: Parameters are rest frequency \citep{mueller2001,lovas2004}, lower level energy $E_l/k$, spectral resolution $\Delta \nu$, and critical density $n_{\rm{crit}}^a$\\
$^a$ Calculated as $n_{\rm{crit}}=\frac{A}{C}$ with Einstein coefficient $A$ and collision rate $C$ at 150\,K.\\
$^b$ No collision rates in LAMBDA database}
\label{lines}                                   
\end{table}                                     

\section{Results}

\subsection{Centimeter continuum emission}
\label{cm_cont}

\begin{figure*}[htb]
  \includegraphics[width=0.99\textwidth]{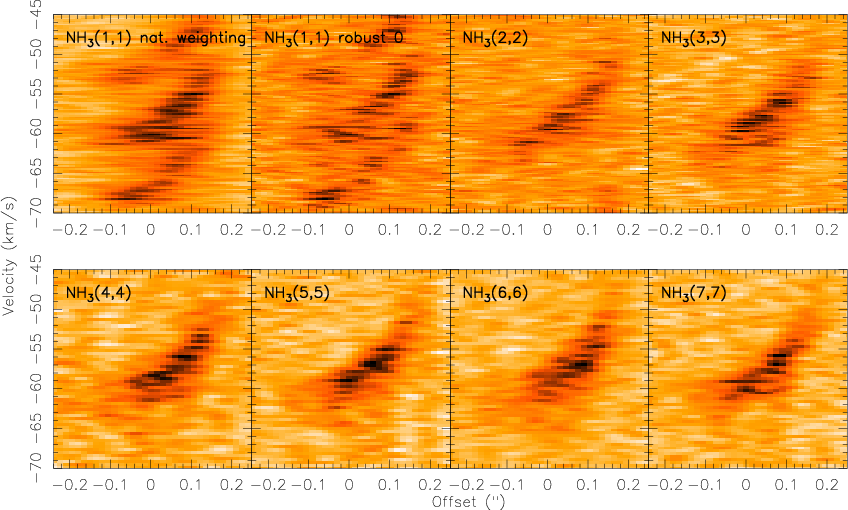}
  \caption{Position-velocity diagrams for the different NH$_3$
    inversion lines as indicated in each panel. The direction of the cut
    is shown in Fig.~\ref{nh3}. The top left and top 2nd panels show
    this cut for natural and robust weighting for the NH$_3$(1,1)
    lines, respectively. All other cuts are carried out for the robust
    weighting case.}
\label{nh3_pv} 
\end{figure*} 

The double-peaked structure of the cm continuum emission presented in
\citet{gaume1995}, \citet{sandell2009}, and \citet{moscadelli2014} can
be interpreted in a two ways: On the one hand, this structure may be
two separate protostellar sources, whereas, on the other hand, the
double-peaked structure could also be part of one underlying jet. Our
new data now allow us to address this question with two approaches:
(a) search for proper motions with a time baseline from December 1992
(\citealt{gaume1995}; observed in the same configuration and
wavelength as our new data) and our new data from July 2015 presented
here. And (b) an analysis of the spectral index based on the broad
bandpass of the new data. Figure \ref{continuum} presents an overlay
of the old and new data as well as a spectral index map. The spectral
index map was derived from all 16 continuum windows between 23.6 and
25.8\,GHz in CASA within the clean task using higher order Taylor
terms (parameter nterm=2) to model the frequency dependence of the sky
emission. The spectral index map is computed as the ratio of the first
two Taylor terms. This task also computes an error map of the spectral
index treating the Taylor-coefficient residuals as errors and
propagating them through the spectral index determination. The
spectral index map in Figure \ref{continuum} is clipped for errors
larger than 0.4.


\begin{table}[htb]
\caption{Continuum source parameters}
\begin{tabular}{lrrr}
\hline
\hline
     & R.A.      & Dec.      & $S_{24.6\rm{GHz}}$ \\
     & (J2000.0) & (J2000.0) & (mJy\,beam$^{-1}$)\\
\hline
cm1  & 23:13:45.368 & 61:28:10.44 & 7.4 \\
cm2  & 23:13:45.377 & 61:28:10.29 & 6.2 \\
\hline                                          
\hline                                          
\end{tabular}                                  
\label{cont}                                   
\end{table}

To search for proper motions, the highest possible positional accuracy
is required. Table \ref{cont} presents the peak positions and peak
fluxes at 24.6\,GHz. When investigating the data for NGC7538IRS1 in detail,
two issues arose. First, the positional accuracy of the
phase calibrator was incorrect by $\Delta\rm{R.A.}\sim 0.01''$ and
$\Delta\rm{Dec.}\sim 0.16''$ \citep{moscadelli2014,goddi2015}. We
corrected for this positional shift after the imaging
process. Furthermore, \citet{moscadelli2009} and
\citet{moscadelli2014} inferred proper motions for the region of
-2.45\,mas\,yr$^{-1}$/-2.45\,mas\,yr$^{-1}$ from CH$_3$OH maser
observations at 12\,GHz. For the 22.58\,yrs time difference between
the two observational epochs, this corresponds to a shift of
$\sim$0.055$''$ in R.A.~ and Dec., respectively. To get the 1992 and
2015 data into the same framework, we shifted the 1992 data according
to these proper motions. Fig.~\ref{continuum} takes both shifts into
account. Furthermore, we show the central positions of the CH$_3$OH
maser groups IRS1a and IRS1b as presented in Fig.~11 of
\citet{moscadelli2014}. Since the maser positions in
\citet{moscadelli2014} are from 2005, we also have to apply the above
proper motion shift to these for comparison. The final CH$_3$OH maser
group positions plotted in this paper are IRS1a: R.A. (J2000.0)
23:13:45.368 and Dec. (J2000.0) 61:28:10.286 and IRS1b: R.A. (J2000.0)
23:13:45.358 and Dec. (J2000.0) 61:28:10.486.

We find that the overall structure of the two central peaks associated
with the two CH$_3$OH maser positions have not moved significantly
within the spatial resolution and uncertainties of the two
observational datasets. The northern source cm1 is elongated in
  northeastern direction, which is already visible in the old data. No
  positional shift can be identified for the northern of the two main
  peaks cm1, whereas the second main source cm2 exhibits a tiny shift
  of $\sim 0.025''$. However, this is less than half of the
  synthesized beam and we refrain from further interpretation of that
  apparently very small shift. The additional emission feature $\sim
0.4''$ to the south appears to have shifted slightly in southeastern
direction. Since we are mainly interested in the two main emission
peaks cm1 and cm2, we do not further analyze the separate southern
structure.
 
\begin{figure*}[htb]
  \includegraphics[width=0.99\textwidth]{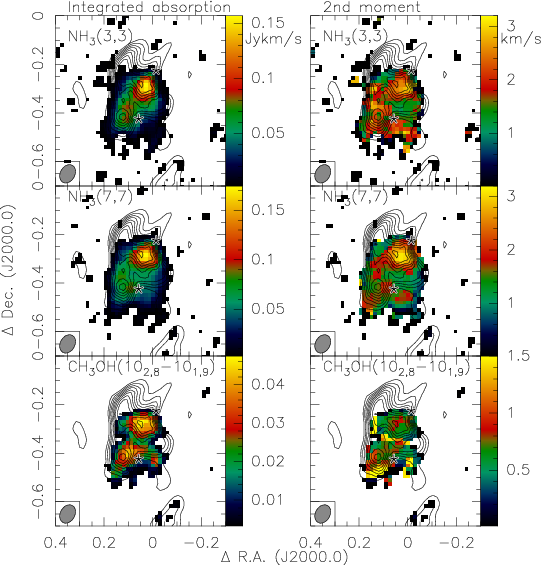}
  \caption{Integrated (left) and 2nd moment (right) maps
    (intensity-weighted velocity dispersion) for two NH$_3$ and one
    CH$_3$OH line as indicated in each panel. The integration ranges
    for NH$_3$(3,3), (7,7) and CH$_3$OH($10_{2,8}-10_{1,9}$) are
    [-68,-50]\,km\,s$^{-1}$, [-65,-50]\,km\,s$^{-1}$ and
    [-65,-53]\,km\,s$^{-1}$, respectively. The contours show in all
    panels the 1.2\,cm continuum emission in levels of 8\% to 98\% of
    the peak emission (7.4\,mJy\,beam$^{-1}$). The two stars indicate
    the CH$_3$OH maser positions by \citet{moscadelli2014}. The
    synthesized beams are shown in the bottom left of each panel.}
\label{moments} 
\end{figure*} 

At the given distance of 2.7\,kpc, our approximate average spatial
resolution element $0.055''$ corresponds to a linear resolution of
$\sim$150\,AU.  Assuming a jet velocity of $\sim$250\,km\,s$^{-1}$
with an inclination angle of 45$^{\circ}$, the 23 years time baseline
between the two observational epochs would still result in proper
motions of $\sim$857\,AU, corresponding in an angular shift of $\sim
0.32''$, which is well resolvable by our observations. Although the inclination
angle is unknown, jets may be even faster
\citep{marti1998,frank2014,guzman2016} and hence these data are
strong evidence that a jet is unlikely the underlying cause for the cm
continuum emission.

\begin{figure*}[htb]
  \includegraphics[width=0.99\textwidth]{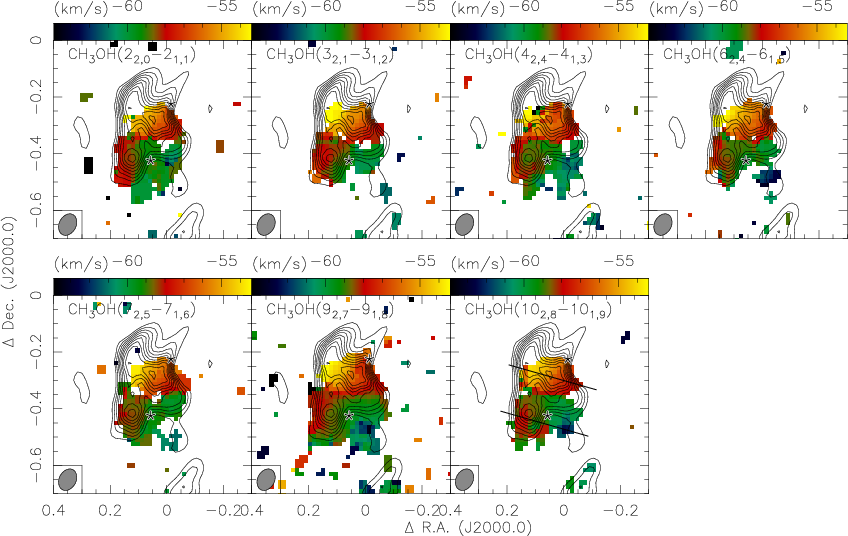}
  \caption{Color-scale presents the 1st moment maps
    (intensity-weighted velocities) of the CH$_3$OH lines as
    shown in each panel. The contours show in all panels the 1.2\,cm
    continuum emission in levels of 8\% to 98\% of the peak emission
    (7.4\,mJy\,beam$^{-1}$). The two stars indicate the CH$_3$OH maser
    positions by \citet{moscadelli2014}. The two lines in the
    bottom right panel indciate the cuts for the position-velocity diagrams shown below.}
\label{ch3oh} 
\end{figure*} 

Regarding the spectral index $S\propto \nu^{\alpha}$, for a
hypercompact H{\sc ii} region $\alpha$ can vary between -0.1 for
optically thin to +2 for optically thick emission. While large H{\sc
  ii} regions are typically in the optically thin regime at the given
frequency around 24\,GHz, a hypercompact H{\sc ii} region such as
NGC7538IRS1 can easily be in the (partly) optically thick regime
(e.g., \citealt{franco2000}). For comparison, while ionized jets
theoretically can cover the same spectral index regime, typical
emission from an ionized jet has rather a spectral index $\alpha$
around +0.6 (e.g., \citealt{reynolds1986,purser2016}). The observed
spectral index $\alpha$ shown in Figure \ref{continuum} varies largely
between 1 and 2. Therefore, the spectral index analysis of
NGC7538IRS1 also indicates that the cm continuum emission is not caused by
an ionized jet but more likely is dominated by a hypercompact H{\sc ii}
region(s).

With the high optical depth indicated by the spectral index,
converting the cm continuum fluxes in the Rayleigh-Jeans limit to
brightness temperatures additionally gives a hint about the temperatures
of the ionized gas. Fig.~\ref{continuum} (right panel) shows that the
brightness temperatures of the inner region vary between
approximately 2000 and 4900\,K. These can be considered as lower
limits for the ionized gas temperatures because the spectral index as
a proxy of the optical depth varies throughout the region.

Combining the multiepoch and multiwavelength analysis above, the
central double-lobe cm continuum emission in NGC7538IRS1 is most likely
emitted by at least two embedded protostars within their associated
hypercompact H{\sc ii} regions.
          
\subsection{NH$_3$ and CH$_3$OH}
\label{molecules}

\begin{figure*}[htb]
  \includegraphics[width=0.99\textwidth]{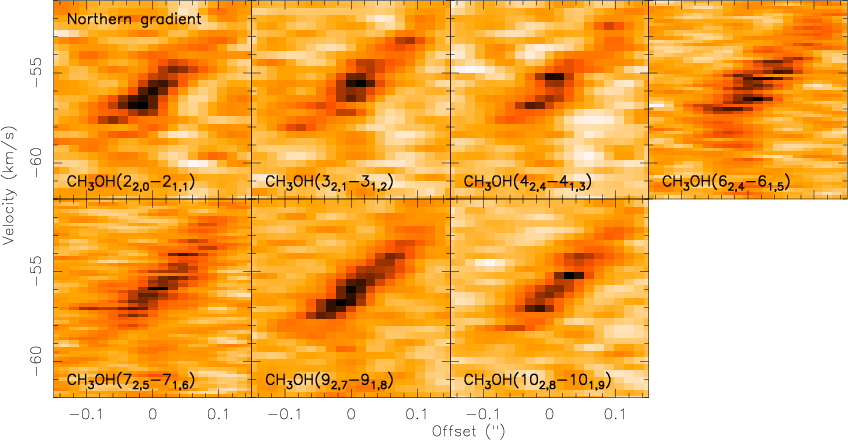}
  \includegraphics[width=0.99\textwidth]{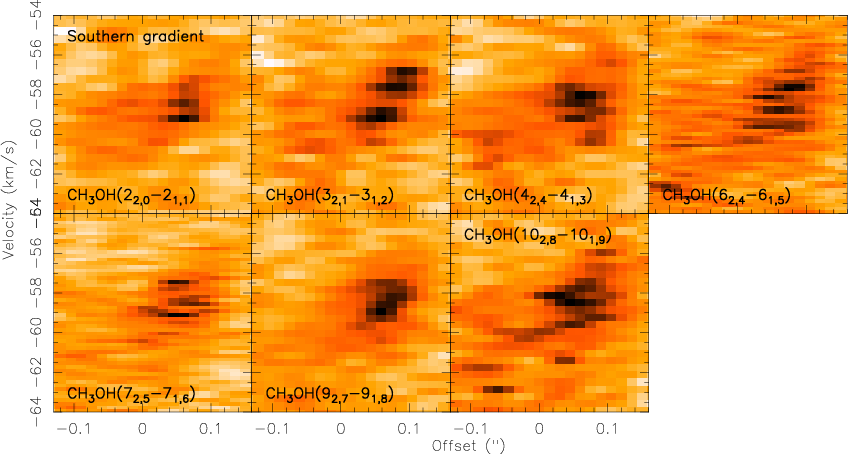}
  \caption{Position-velocity diagrams for the different CH$_3$OH lines
    as indicated in each panel. The top seven panels belong to the
    northern cut and the bottom seven panels to the southern cut shown
    in Fig.~\ref{ch3oh}. All cuts are performed for the robust weighting 0
    case.}
\label{ch3oh_pv} 
\end{figure*}

At this high spatial resolution and with the given very strong
continuum emission, all molecular line features are only observed in
absorption against the continuum sources. Figures \ref{spectra_nh3}
and \ref{spectra_ch3oh} show example spectra of all NH$_3$ and
CH$_3$OH lines extracted toward the northern cm continuum peak
position. For NH$_3$ the hyperfine structure is detected for all
lines. The integrated, 1st and 2nd moment maps and the
position-velocity diagrams discussed below (Figs.~\ref{nh3} to
\ref{ch3oh_pv}) were created after inverting the data (simply
multiplying by -1) because the corresponding algorithms in CASA only
work on positive data. This inversion does not affect the kinematic
signatures at all. The excitation temperatures $E_l/k$ and the
critical densities calculated as $n_{\rm{crit}}=A/C$ (with the
Einstein coefficient $A$ and the collision rate $C$ from the LAMBDA
database; \citealt{schoeier2005}) are given in Table \ref{lines} as
well. While NH$_3$ covers an excitation range between 22 and 537\,K,
this is slightly smaller for CH$_3$OH between 28 and 149\,K. However,
while the critical densities $n_{\rm{crit}}$ for NH$_3$ are around
2000\,cm$^{-3}$, they are more than an order of magnitude larger for
CH$_3$OH around a few times $10^4$\,cm$^{-3}$. In addition to this,
the chosen $J_2 - J_1$ 25\,GHz transitions of CH$_3$OH have been found
to emit as masers in several high-mass star-forming regions (e.g.,
\citealt{menten1986b,voronkov2007,brogan2012}). These masers are
likely collisionly excited (e.g., \citealt{sobolev1983}) and thus form
its own subgroup of methanol Class I maser (e.g.,
\citealt{leurini2016}). Compared to other types of Class I masers,
they need higher gas volume densities for the population inversion to
occur ($n > 10^6$\,cm${^-3}$; e.g.,
\citealt{sobolev1998,leurini2016}). The fact that we see the 25\,GHz
transitions in absorption might be an indication that the majority of
the methanol in NGC 7538 IRS1 relevant for our absorption is residing
in somewhat lower density gas.

Fig.~\ref{nh3} presents the 1st moment maps (intensity-weighted
velocities) of the NH$_3$ inversion lines from (1,1) to (7,7). The
first two maps of the (1,1) transition are produced with different
weighting schemes (natural weighting and robust weighting 0)
recovering different spatial structures. While the naturally weighted
NH$_3$(1,1) shows a bit more extended emission, one sees a
larger scale velocity gradient approximately in east-west
direction. In comparison to that, the higher resolution (robust
weighting 0) image rather reveals a velocity gradient in
northeast-southwest direction, which is consistent with previous the findings by
\citet{beuther2012c,beuther2013b}, \citet{moscadelli2014}, and
\citet{goddi2015}. Since we are mainly interested in the kinematics of
the innermost central sources we concentrate in the following on the
higher resolution data. All other transitions are also presented in
this higher resolution imaging mode (robust weighting 0).

Interestingly, all seven NH$_3$ inversion lines with excitation levels
$E_l/k$ between 22 and 537\,K (Table \ref{lines}) exhibit almost the
same kinematic structure of the central core with one velocity
gradient in approximately northeast-southwest direction, but without
any differentiation between the two continuum sources seen in the
overlayed contours (Fig.~\ref{nh3}). These NH$_3$ velocity structures
appear similar to the previous findings for the rotational structure
of the rotating envelope by \citet{beuther2013b} and
\citet{goddi2015}. Performing spectral cuts across the main velocity
gradient direction (NH$_3$(7,7) panel in Fig.~\ref{nh3} indicates the
exact orientation), Figure \ref{nh3_pv} presents the corresponding
position-velocity diagrams. The two panels corresponding to the (1,1)
inversion lines exhibit three features because of the
close-in frequency-space hyperfine structure. The other six lines only show
the central strong hyperfine component. This velocity gradient is
almost linear across the source. It does not show any hint of
Keplerian motion but resembles more a solid-body rotation diagram.

For comparison, Fig.~\ref{moments} presents also the integrated
absorption and the 2nd moment maps (intensity-weighted velocity
dispersion) for two selected NH$_3$ and one CH$_3$OH lines. The
spatial structure of these integrated and velocity dispersion maps
does not reflect the overall velocity gradient seen in NH$_3$, but all
maps are double-peaked toward the two cm continuum peak
positions. This shows that the largest gas column densities and
strongest line broadening are indeed associated with the two main
protostellar condensations. The fact that NH$_3$ exhibits two centers
of line broadening, in spite of only a single larger scale velocity
gradient, indicates the potential existence of two smaller scale
embedded rotating structures.

The important new information now stems from the thermal CH$_3$OH
absorption data. Similar to NH$_3$, for CH$_3$OH we also present the 1st
moment maps and position-velocity diagrams in Figures \ref{ch3oh} and
\ref{ch3oh_pv}. While the CH$_3$OH 1st moment maps also exhibit the
general trend of velocities from the northeast to the southwest, the
data show a clear structural change between the northern and southern
continuum source. With these data one can depict for all lines with
excitation levels between 28 and 149\,K (Table \ref{lines}) one
velocity gradient across the northern continuum source and one
velocity gradient across the southern continuum source. These two
velocity gradients are almost parallel in the east-northeast to
west-southwest direction. While we identify these velocity gradients
in thermal CH$_3$OH absorption, the
previously studied CH$_3$OH class II masers also show velocity
gradients approximately in the east-west direction
\citep{moscadelli2014}. Although the angles derived from the maser and
thermal absorption are not exactly the same, they are both
approximately in east-west direction and both have the same
orientation with respect to the blue- and redshifted structure. Hence,
while the maser and thermal emission and absorption trace different
spatial scales, both appear to stem from the same rotating structures.

Figures \ref{ch3oh_pv} presents the corresponding position-velocity
cuts along the two axes outlined in the bottom right panel of
Fig.~\ref{ch3oh}. For both regions we identify clear velocity
gradients across the sources, however, in both cases again without any
Keplerian signature. The underlying physical reasons for these
kinematic signatures are discussed in section \ref{kepler}.

\begin{figure}[htb]
  \includegraphics[width=0.49\textwidth]{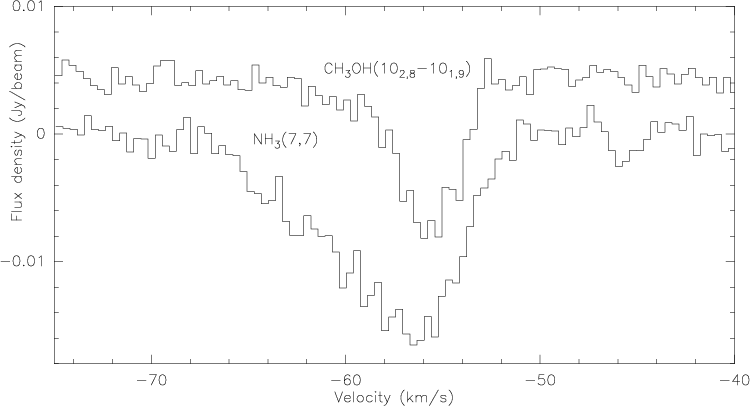}
  \caption{Example absorption spectra of NH$_3$(7,7) and
    CH$_3$OH($10_{2,8}-10_{1,9}$) taken toward the northern cm
    continuum peak position. The CH$_3$OH spectrum is shifted slightly
    to positive flux density values for better presentation.}
\label{abs} 
\end{figure}

While the measured velocity dispersion of the two NH$_3$ lines varies
only a bit, the velocity dispersion of CH$_3$OH is considerably
narrower (Fig.~\ref{moments}). Only the overlap region between the two
continuum peaks shows a larger velocity dispersion, but this can be
attributed to beam smearing effects between the two peak
positions. Inspecting individual absorption spectra against the main
northern continuum peak position (Fig.~\ref{abs}), the spectral
profiles show why the measured line widths in NH$_3$ and CH$_3$OH are
different. While the peak and redshifted side of the absorption
spectra are similar, NH$_3$ shows a pronounced blueshifted wing. In
absorption spectra that is clear sign for outflowing gas. Because the
critical density of NH$_3$ is an order of magnitude lower than that of
CH$_3$OH (Table \ref{lines}), it appears that NH$_3$ also traces
outflowing gas of within the envelope whereas the CH$_3$OH signatures
are more dominated by the rotating disk-like structures.

\section{Discussion}

\subsection{Fragmentation and multiplicity}

The high-mass star-forming region NGC7538IRS1 is intriguing because it
does not show significant fragmentation signatures in the cold dust
and gas emission at (sub)mm wavelengths. At $\sim 0.3$ resolution and
1.3\,mm wavelength, \citet{beuther2012c} still identified only a
single source, whereas then at $\sim 0.2''$ resolution and 843\,$\mu$m
wavelength first fragmentation signatures of the innermost rotating
structure could be identified \citep{beuther2013b}. Our new
(non-)proper motion analysis, spectral index study and the kinematic
signatures in CH$_3$OH clearly confirm that at least two very young
protostars are embedded within the innermost core. The elongation in
northeastern direction of the northern source cm1 may harbor further
subsources. However, this elongation may also be due to an underlying
ionized disk-like structure since it is approximately aligned with the
CH$_3$OH velocity gradient. The projected separation of the two main
protostars cm1 and cm2 is $\sim 0.16''$ or $\sim 430$\,AU. Since the
two sources are embedded in a large-scale rotating structure
(Fig.~\ref{nh3}), they are most likely a bound binary system.

While the rotation axis of the disk-like structures around these two
protostars are almost parallel (Fig.~\ref{ch3oh}), the rotation axis
of the surrounding gas envelope is inclined to the disk-like
structures (Fig.~\ref{nh3}). As the NH$_3$ velocity gradient is
approximately 41$^{\circ}$ east from north and the CH$_3$OH velocity
gradient is approximately 74$^{\circ}$ east from north, the relative
inclination between the two axis is 33$^{\circ}$. Why are the axes of
the disk-like structure around the two protostars and the surrounding
envelope not better aligned? In turbulent molecular clouds one can
easily consider a collapse scenario in which the different collapsing
shells within the envelope have varying angular momentum distributions
already at the beginning of the collapse. Therefore, gas that falls
earlier and deeper into the gravitational potential well of the
forming cluster can have different angular momentum vectors than the
remnant envelope that may still feed the inner disk-like
entities. Therefore, in this scenario, misalignment between axes on
different spatial scales can qualitatively be well understood. Similar
results were recently obtained by \citet{kraus2017} with VLTI
observations toward the high-mass binary system IRAS\,17216-3801.

\subsection{(Non-)Keplerian motions}
\label{kepler}

Where is the high-velocity gas one would expect from an embedded
Keplerian disk? Two main options exist as potential answers. First, it
can be a spatial resolution issue where the Keplerian velocities are
hidden below our angular resolution. With a spatial resolution of
$\sim 0.07''$, corresponding to a linear resolution of $\sim$190\,AU
that would limit the size of potential Keplerian structure to below
that scale. However, in that picture the observations would in
principle still see the high-velocity gas, just smear it out over the
beam size. Hence, some remaining high-velocity gas could even be
observable at this spatial resolution.

Second, it may also be a physical effect because we recall that the
presented data are absorption line observations. Hence, we only
observe the gas in front of the hypercompact H{\sc ii}, and we
explicitly miss the innermost ionized gas within the hypercompact
H{\sc ii} region. Considering the scenario outlined first by
\citet{keto2002a,keto2003} in which the accretion flow changes from
molecular to ionized form within the inner hypercompact H{\sc ii}
region, we would be missing the highest velocity gas by such molecular
observations anyway. Estimating the source size of the central
hypercompact H{\sc ii} regions is difficult in such a crowded
area. However, based on Fig.~\ref{continuum}, we can estimate the
projected size to $\sim 0.1''-0.2''$. Assuming a spherical source
structure and that only the front half is part of our observations, we
are missing $\sim 0.05''-0.1''$ along the line of sight in our
molecular gas data. That corresponds to linear scales of $\sim
135-270$\,AU. Therefore, in both scenarios, we cannot resolve the
central highest velocity gas structures. However, while the first
simple spatial resolution argument would still ``see'' the
high-velocity gas -- just smeared out over the beam size -- the second
physical argument of an inner ionized H{\sc ii} does not allow us to
see that high-velocity gas at all in such molecular absorption line
data.

What scales are predicted by simulations for Keplerian structures
around high-mass accretion disks? For example, \citet{krumholz2007a}
present position-velocity diagrams for simulations around a forming
high-mass star (8.3\,M$_{\odot}$ at the presented time step) where the
Keplerian signatures could be visible at least out to radii of
250\,AU. \citet{kuiper2011} show the time evolution of Keplerian
structures around forming massive stars and the Keplerian size
increases with time. In the model of a collapsing 60\,M$_{\odot}$
core, the Keplerian structure grows from below 100\,AU at times
earlier than $10^4$\,yrs to more than 1000\,AU after $5\times
10^4$\,yrs. While these are only simulated individual cases studies,
they already outline the range of potential Keplerian disk sizes. With
the observed infall (e.g., \citealt{beuther2013b}), NGC7538IRS1 should
still be at a comparably early evolutionary stage, hence small disk
sizes are possible. Furthermore, our data clearly show the multiple
structure of the region, which can truncate disks even further. 

These observations clearly outline the complicated nature of studying
high-mass accretion disks. On the one hand, extremely high spatial
resolution at sub-$0.1''$ is required. However, that resolution is achievable now
with observations at cm wavelength at the VLA, such as those presented
here, or new observations with the Atacama Large Millimeter Array
(ALMA) that can reach even higher resolution; however, this target
NGC7538IRS1 is too far north and not accessible with ALMA.

On the other hand, it is also crucial to identify the right sources
and disk tracers. If we are dealing with hypercompact H{\sc ii}
regions, ionized tracers such as radio recombination lines could be very
useful (e.g., \citealt{keto2008,keto2008b,klaassen2009}). For
NGC7538IRS1, we did observe such recombination lines simultaneously,
however, the sensitivity was insufficient for further
analysis. Furthermore, as shown in \citet{keto2008}, where they
present radio recombination line data toward NGC7538IRS1 at 22 and
43\,GHz, these lines at cm wavelengths are typically very broad. A
significant amount of the line width at these wavelengths is caused by
thermal and pressure broadening and disentangling kinematic
signatures from these components is not trivial. Going to (sub)mm
wavelengths may improve the situation because there at least the
pressure broadening is significantly reduced. Therefore, it may be
more promising to select sources at even earlier evolutionary phases
where no hypercompact H{\sc ii} region has formed yet;  one hence
can study the kinematics at much smaller scales in the molecular form.

\section{Conclusions}
\label{conclusion}

Resolving the famous high-mass star-forming region NGC7538IRS1 at the
highest spatial resolution possible at cm wavelengths with the VLA
($0.06''\times 0.05''$ corresponding to $\sim$150\,AU) reveals several
new insights into the physics of this archetypical high-mass
star-forming region. Comparing the new data to previous epoch
observations from $\sim$23\,yrs ago, no proper motions can be
identified. In combination with a high spectral index largely varying
between 1 and 2, we infer that the cm continuum emission does not stem
from an underlying jet, but that it is rather dominated by two
hypercompact H{\sc ii} regions that are likely formed by two separate
high-mass protostars. Based on the kinematics, these protostars appear to form a
bound system within a circumbinary envelope.

The CH$_3$OH and NH$_3$ spectral line data reveal different velocity
structures in absorption against the strong continuum emission. The
thermal CH$_3$OH data show two velocity gradients across the two
continuum sources, indicating the existence of two embedded disk-like
structures. The approximate orientation and velocity structure of
these thermal CH$_3$OH measurements agree well with the much higher
resolution CH$_3$OH maser data \citep{moscadelli2014}. While the two
disk-like structures are almost parallel, the NH$_3$ data trace a
rotating circumbinary envelope that is inclined to the two disk-like
structures by $\sim$33$^{\circ}$. Such variations in rotation axis
between envelope- and disk-structures can be caused by varying initial
angular momentum distribution in the natal, turbulent molecular cloud.

The fact that we do not identify Keplerian signatures in the
disk-tracing CH$_3$OH data is mostly caused by the nature of this
molecular absorption line data, which do not trace the innermost gas
that is ionized already. A closer investigation of the kinematics to
the center will require recombination line observations, best
conducted at (sub)mm wavelengths where the pressure broadening of the
line becomes negligible.

\begin{acknowledgements} 
  We like to thank Luca Moscadelli and Ciriaco Goddi for providing the
  cm continuum data from 1992 and for giving the details about the
  positional uncertainties of the calibrator and the proper motions of
  the region. Furthermore, we thank Stella Offner for an interesting
  discussion about angular momentum vector distributions and their
  variations in turbulent cores. We would also like to thank the
  referee for a constructive and helpful report. This paper makes use
  of data from the Karl G.~Jansky Very Large Array. The National Radio
  Astronomy Observatory is a facility of the National Science
  Foundation operated under cooperative agreement by Associated
  Universities, Inc.  HB acknowledges support from the European
  Research Council under the Horizon 2020 Framework Program via the
  ERC Consolidator Grant CSF-648505.

\end{acknowledgements}


\end{document}